\documentclass[prl,twocolumn,aps,showpacs,superscriptaddress]{revtex4-1}
\usepackage{graphics}
\usepackage{epsfig}
\usepackage{amsmath, bm}
\usepackage{amssymb}
\usepackage{color}
\usepackage{soul}
\usepackage{hyperref}
\graphicspath{{Figures/}}
\newcommand{\rom}[1]{\uppercase\expandafter{\romannumeral #1\relax}}

\begin{document}
\newcommand{\beq}{\begin{equation}}
\newcommand{\eeq}{\end{equation}}

\title{Multi-Frequency Acoustic Steering in Rationally Pruned Disordered Networks}

\author{Anastasiia O. Krushynska}
\affiliation{Engineering and Technology Institute Groningen (ENTEG), Faculty of Science and Engineering, University of Groningen, Nijenborgh 4, 9747 AG Groningen, The Netherlands}
\author{Martin van Hecke}
\affiliation{AMOLF, Science Park 104, 1098 XG Amsterdam, The Netherlands}\affiliation{Huygens-Kamerlingh Onnes Lab, Leiden University, PObox 9504, 2300 RA Leiden, The Netherlands}

\date{\today}

\begin{abstract} 
The frequency of a wave is a crucial parameter to understanding its propagation to a structured medium. 
While metamaterials can be designed to focus or steer deformations or waves toward a specific location, wave steering towards multiple targeted locations at different frequencies remains challenging.
Here we show that pruning of random elastic networks allows realizing such multi-objective, multi-frequency acoustic steering. Our study opens up a 
viable route to the rational design of multi-frequency acoustic metamaterials.
\end{abstract}


\maketitle

The rational design of metamaterials, also known as the inverse problem, is challenging.
A range of approaches
\cite{goodrich15, combinatorial16, SerraGarcia18,  natphys_mechanisms}
now allows acoustic and mechanical  metamaterials to achieve
acoustic and static cloaking 
\cite{acousticcloak08, wegener_unfeelability14},
non-reciprocal propagation
\cite{Alu_waves14, AluCoulaisNature17},
topological protection
\cite{Vitelli14, Huber18, Zeb17, Rocks21, Kane14, Meeussen16},
strain steering
\cite{rocks17, Meeussen20}
as well as switchable
\cite{mullin2007, shim12},
programmable
\cite{silverberg2014, florijn2014},
and shape-morphing behavior
\cite{silverberg2015, combinatorial16, waitukaitis2018}.
Specifically, acoustic metamaterials can steer waves along predefined paths
\cite{Cummer16, Casadei12, Fleury16, Rupin14, Fleury18},
allowing applications including enhanced imaging  \cite{Zhu11},
focusing
\cite{Guenneau07},
data processing
\cite{Raney16}
and wave based computing
\cite{Silva14, Fleury21}.
These strategies are generally based on lattices. One strategy works by
predefined wave guiding paths that connect input and output nodes, realized by modifying unit cells along a given path, leading to broad band steering of acoustic energy 
\cite{Casadei,Cicek, Yuan}.
Another strategy leverages topology optimization of the unit cells to vary the wave transmission with frequency over a limited range in amplitudes \cite{Rong}. 
However, it is an open question if and how one can realize multi-frequency wave steering with multiple-targets --- which likely cannot work if all paths have to be predefined.

\begin{figure}[t!]
\begin{tabular}{c}
\includegraphics[scale=0.75,bb = 30 30 190 200,clip]{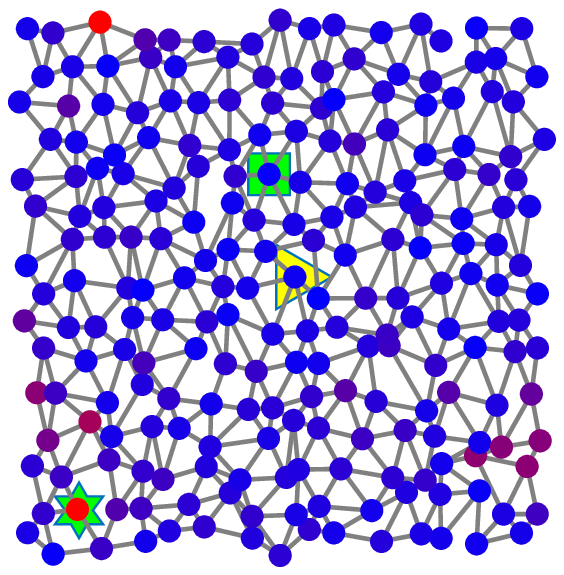}
\includegraphics[scale=0.75,bb = 30 30 190 200,clip]{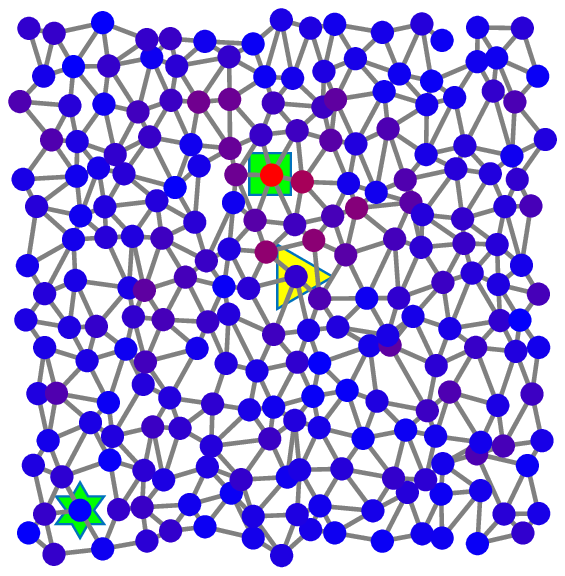}\\
\end{tabular}
\caption{(color online) Exciting the input node (triangle) of a rationally pruned network (43 springs pruned)
with input frequency (a) $f_1=1$ or (b) $f_2=1.2$
directs the vibration energy towards output node 1 (square) or output node 2 (star). Colors of the node indicate their normalized amplitude (blue: 0, red: 1); in this example we achieve
amplitude contrast $c=50$.
}
\label{fig:1}
\end{figure}

It was recently shown that starting from a fully disordered network structure, rational pruning strategies can be employed to tune the {\em static} elastic properties
of disordered metamaterials over a wide range~\citep{goodrich15}.
Such `tuning by pruning' strategies are straightforward to implement and allow to
explore a much larger design space than is possible with lattice-like metamaterials
\citep{rocks17}.
In particular, this strategy has been explored to achieve targeted propagation of deformations, resulting in ``allosteric'' metamaterials where static forcing at one location yields a predefined response at a distant location
\cite{rocks17, Wyart17},
or to control flow in disordered transport networks
\cite{rocks2019, rocks2020}.
Strikingly, these functionalities do not require to predefine a path, but are
rather collective properties --- the resulting geometries do not feature
discernible paths connecting input and output nodes.
So far, only quasistatic properties have been considered. It is thus a natural question to ask
if pruning strategies are suitable for wave steering in acoustic metamaterials. The driving frequency provides a natural control parameter for acoustic problems with no static counterpart. This allows us to ask whether we can achieve multiple target functionalities at different frequencies in the same geometry.

\begin{figure*}[t!]
\begin{tabular}{cc}
\includegraphics[width=1\columnwidth]
{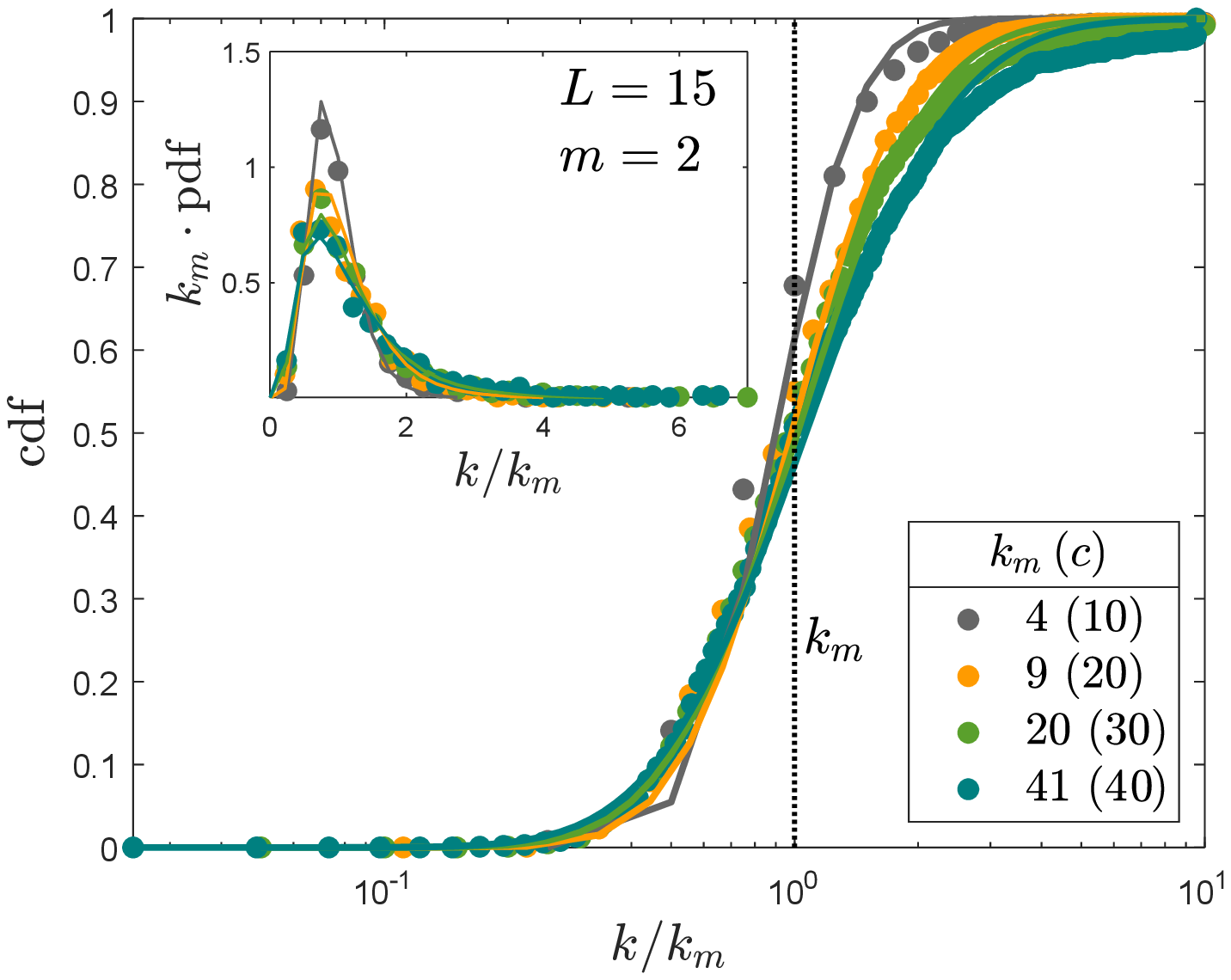}
&
\includegraphics[width= 1.05\columnwidth]
{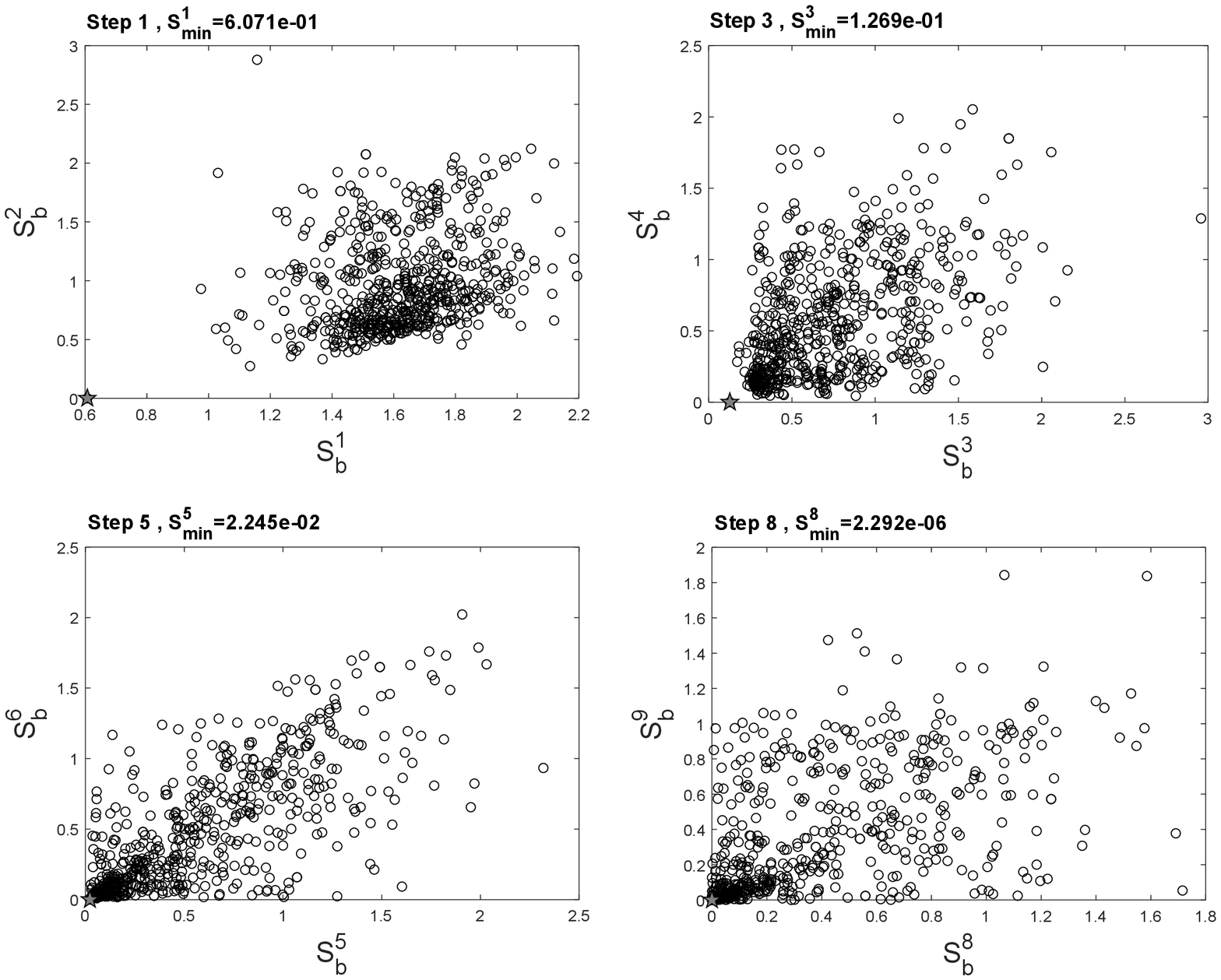}
\end{tabular}
\caption{(color online)
(a) The cumulative distribution function (cdf) and probability density function (pdf, inset) for the number of pruned springs $k$ scaled by the media $k_m$ for a range of values of the contrast $c$ (1000 systems, $L\!=\!15$, $m\!=\!2$, target patterns 10 for $f_1\!=\!1$ and 01 for $f_2\!=\!1.2$).  The data follows a log-normal distribution (solid).
(b) Representative scatter plots of subsequently removed bonds
$S_b^{k+1}$ vs $S_b^{k}$ during pruning (parameters as in (a) and $c\!=\!30$). {The points corresponding to pruned bonds are indicated by the stars.}
}
\label{fig:2}
\end{figure*}

Here we investigate multi-target acoustic steering by rational pruning of disordered networks. To do so, we inject vibrational energy into a disordered lattice at a given location, and aim to achieve
specific patterns of vibrations at a number of output nodes for a number of distinct frequencies --- in other words, we use the excitation frequency to steer acoustic energy to different patterns of output locations (Fig.~\ref{fig:1}). We find that a simple greedy pruning algorithm is able to do so, including for more than two target nodes or frequencies. To understand the capacity of random networks to embed acoustic steering, we study the nature of the pruning process, as well as the
statistics of pruning as function of the system size, number of output nodes and target patterns. We find that the linear size of the network controls the
typical number of pruned bonds required to reach a given target functionality, which suggest that wave pruning is sub-extensive.
Together, our work shows that pruning strategies on disordered networks are a simple, flexible and effective strategy
to realize multiple-target wave steering, thus opening up new possibilities for advanced acoustic multiplexing,
data processing and computing
\cite{Silva14, Fleury21}.

{\em Spring Networks.---} 
We study linear vibrations of spring networks driven harmonically at a single input node with amplitude 1 and horizontal polarization.
Full networks are created by randomly perturbing the $L_x \times L_y$ nodes of a triangular lattice by up to 35\%. These networks are then sequentially pruned so that different frequencies yield specific on/off amplitude target patterns at preselected output nodes (Fig.~\ref{fig:1}).
We take periodic boundary conditions,
unit mass and spring constants, and the unit of time as $\sqrt{1/2 \pi}$. We have studied the inverse participation ratio of our modes as function of frequency, which indicates that the modes become quasi-localized for $f\gtrsim 2.5$ and become plane waves at very low frequencies (Supplemental Information). In the remainder, we restrict our driving frequencies
between $0.5$ and $2$, which ensures that the vibration modes are disordered yet not localized.

{\em Pruning.---}
We prune full networks by removing $k$ bonds, so that $n$ preselected frequencies $f_j$ yield $n$ specific on/off amplitude target patterns at $m$ preselected output nodes; for example,
aiming for target patterns `010' and `101' corresponds to
$m\!=\!3$ and $n\!=\!2$. We note that typical node amplitudes $\tilde{A}_i$ vary strongly with frequency due to resonances, and define normalized node amplitudes: $A_i:= \tilde{A}_i / \sqrt{\Sigma_i \tilde{A}_i^2}$. Moreover, we note that achieving strictly zero amplitudes is not feasible, and instead aim for high and low target amplitudes which satisfy $A\!>\!A_1$ and $A\!<\!A_0$ respectively, with contrast $c\!:=\!A_1/A_0$. We fix $A_0=0.1$, so that $A_1 = c/10$, and large contrasts $>10$ imply wave suppression at the low-amplitude nodes, and amplification at the large-amplitude nodes.
We define a piece-wise linear function
$f$ which maps large and small amplitude to zero and one, with linear interpolation in between:
\begin{equation}
f(A_i)=
\begin{cases}
1, & \text{if } A_i \geq A_1~, \\
\frac{A_i-A_0}{A_1-A_0}, & \text{if }A_0<{}A_i<{}A_1~,\\
0, & \text{if } A_i \leq A_0 ~,\\
\end{cases}
\end{equation}
and define a cost
function $S$ which reaches zero when the amplitudes match the target pattern:
\begin{equation}
S = \sum\limits_{i,j} (A^t_{i,j} - f(A_{i,j}))^2~.
\end{equation}
Here $A^t_{i,j}$ are the binary target amplitudes at node $i$ and frequency $j$; for example, steering waves between two frequencies would correspond to taking both
$A^t_{1,1}$ and $A^t_{2,2}$ equal to one, and
$A^t_{1,2}$ and $A^t_{2,1}$ equal to zero.

We use greedy pruning to minimize $S$: at each step in the algorithm we remove the bond that yields the minimal value of $S$.
We continue this process for $k$ steps, until either
the cost function $S$ becomes numerically zero and a satisfactory network is found, or $k$ exceeds $L_x L_y$; we never see convergence when more bonds are removed, and in any case, rigidity is lost then.

{\em Phenomenology.---} We illustrate our approach
for a system of $15^2$ nodes with $m=2$ target nodes, $n=2$ target frequencies, $C=50$, and
two ``switching'' target patterns  (target `01' at $f_1$ and `10' and $f_2$). For these particular parameters, our greedy pruning strategy is almost always successful in producing a geometry (Fig.~\ref{fig:1}).
We stress that this example demonstrates a strong contrast ($c=50$)
by pruning only seven percent of the springs,
showing the feasibility of pruning to achieve acoustic steering.
Finally, we note that the spatial patterns of pruned bonds reveal no
(wave guiding) path connecting the input and output nodes and no discernible
correlations to, e.g., target nodes (Supplemental Information).

We have studied the statistics for the required number of pruned bonds, $k$, and find that its distribution has a long tail for large $k$ and is well fitted with a log-normal distribution (Fig.~\ref{fig:2}a), irrespective of contrast, system size and number of target nodes (Supplemental Information). For more challenging targets, not all runs converge. We therefore
characterize the distribution of $k$ by its median $k_m$, which remains well-defined as long as more than half of the runs converge.

There is a major difference between our acoustic pruning and previous pruning strategies for static properties. In the latter case, the target function can be expressed as a sum over bonds so that bonds can be removed independently
\citep{goodrich15, rocks17}.
In contrast, in our system bond removal modifies the eigenmodes, which we therefore recalculate after each pruning step.
We have probed potential correlations between subsequent bond removals by investigating scatter plots of $S_b^k$, the value of $S$ after removing bond $b$ in generation $k$, and $S_b^{k+1}$.
First, we notice that the convergence of $S$ is not strictly monotonic, in particular after many steps where the differences in $S$ over subsequent generations are small (Fig.~\ref{fig:2}b).
Second, we find that correlations between
$S_b^k$ and $S_b^{k+1}$ intermittently build up and decay (Fig.~\ref{fig:2}b and Supplemental Information).
While removing multiple bonds in one step may be feasible
during correlated episodes, the frequent and unpredictable presence of uncorrelated episodes shows that tuning the vibrational properties
requires costly recalculations of the spectrum at each step.

\begin{figure}[t!]
\begin{tabular}{c}
\includegraphics[width= 0.95\columnwidth]
{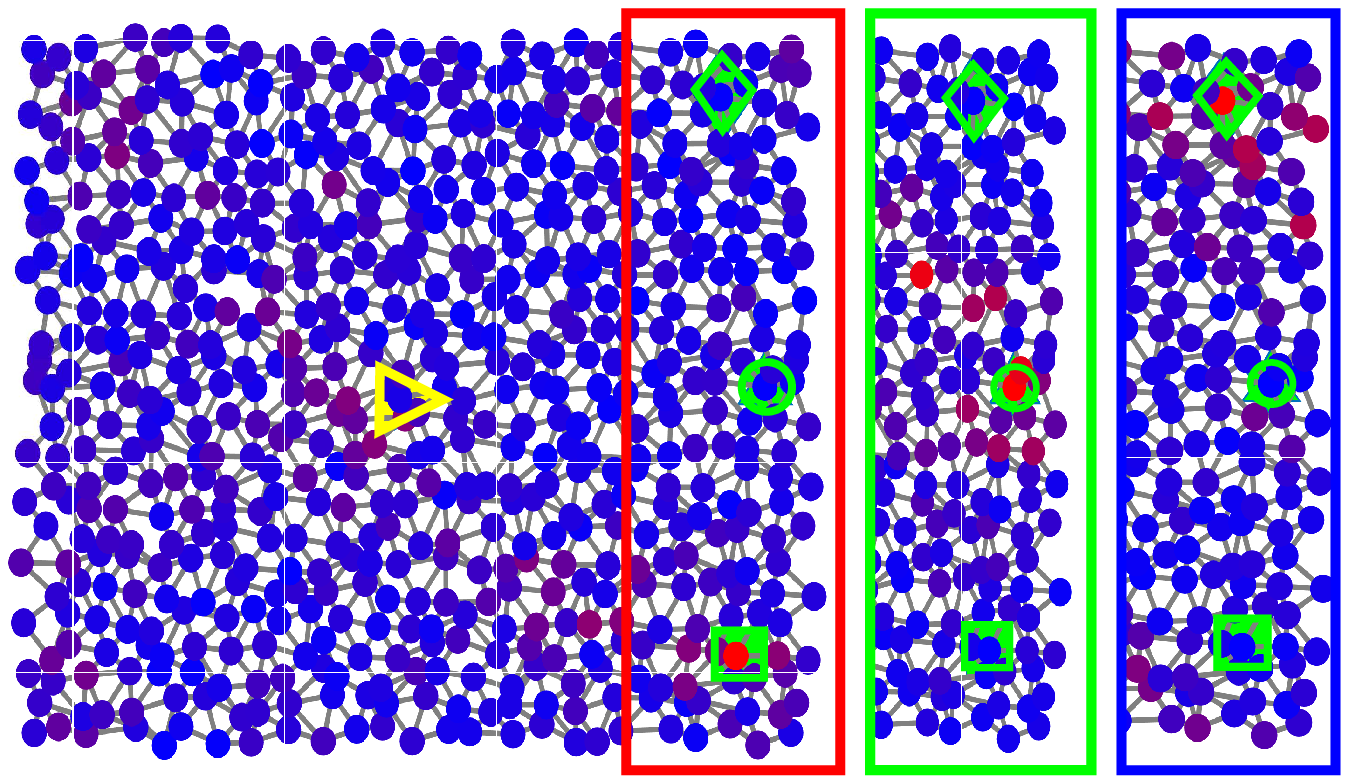}\\
(a)\\
\includegraphics[width= 0.95\columnwidth]
{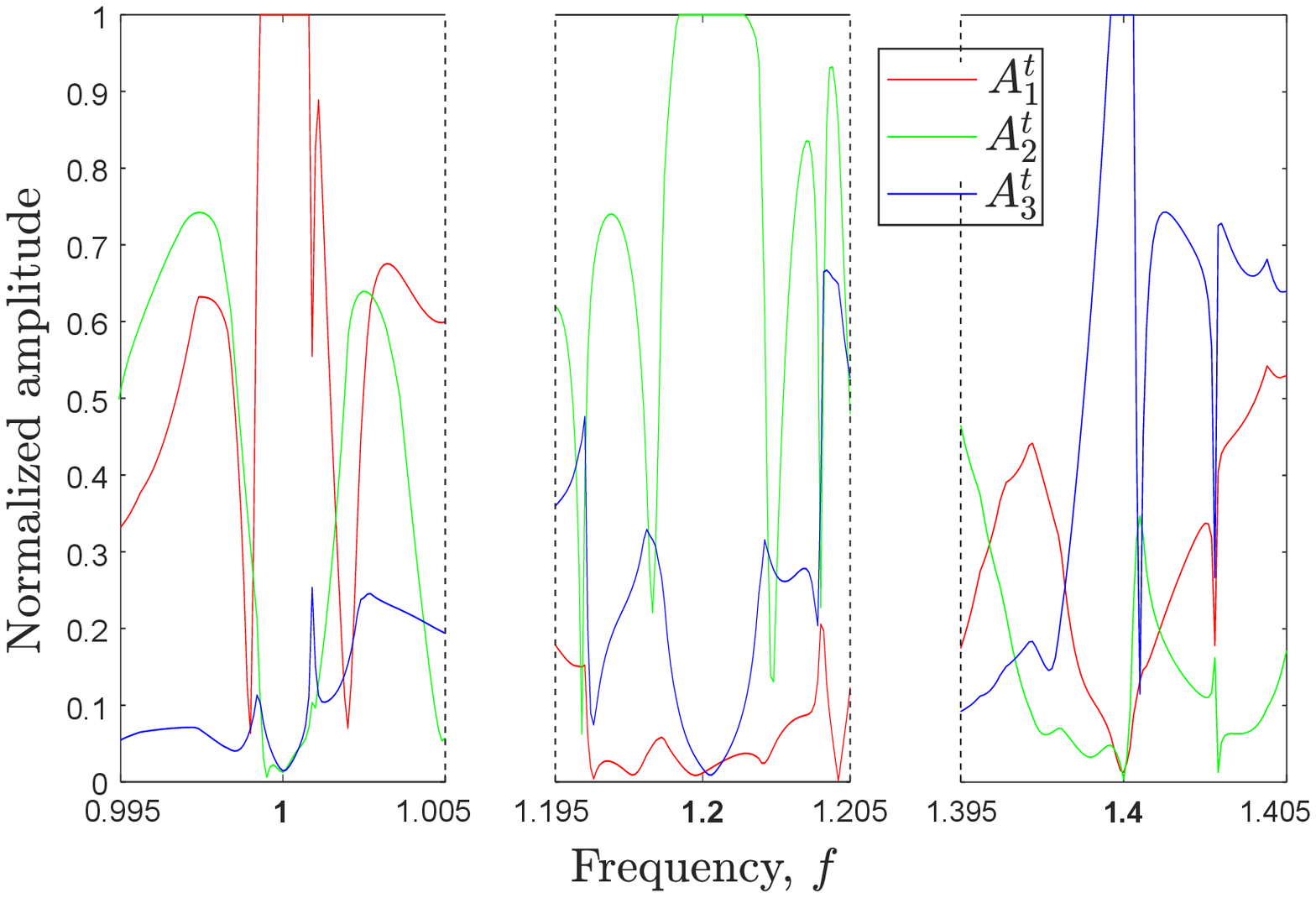}\\
(b)
\end{tabular}
\caption{(color online) (a) Acoustic steering in a system with $L=25$ and $m=3$ with mixed patterns `100' at $f_1=1$, `010' at $f_2=1.2$, `001' at $f_3=1.4$. To achieve contrast $c=50$, we pruned 117 bonds. (b) The target amplitudes preserve their values in a close vicinity of the excitation frequency.}
	\label{fig:3}
\end{figure}

{\em Acoustic steering with multiple targets.---}
We have also explored if we can steer sounds waves to three separate nodes at three different frequencies.
We consider a large system ${L=25}$ and define ${m=3}$ with target patterns: `100' at $f_1=1$,  `010' at $f_1=1.2$, and `001' at $f_1=1.4$. We find we can reach a contrast $c=50$ by only pruning between three and 15 percent of the bonds (an exemplary pruned network in Fig.~\ref{fig:3}a).
While we fix the locations of target nodes to the right-hand side of the network, we find that for other locations we can reach similarly good contrasts (Supplemental Information).
We note that the amplitudes at non-target nodes again appear random and do not reveal specific  paths.
Plotting the response at the three nodes as function of frequency, we see that these are complex,
yet reaching large or small values near the target frequencies (Fig.~\ref{fig:3}b).
This data reveals that the steering is robust to small variations in the excitation frequencies, as the variations of target amplitude near the target frequencies is relative slow.

\begin{figure}[ht!]
\begin{tabular}{c}
\includegraphics[width= \columnwidth]
{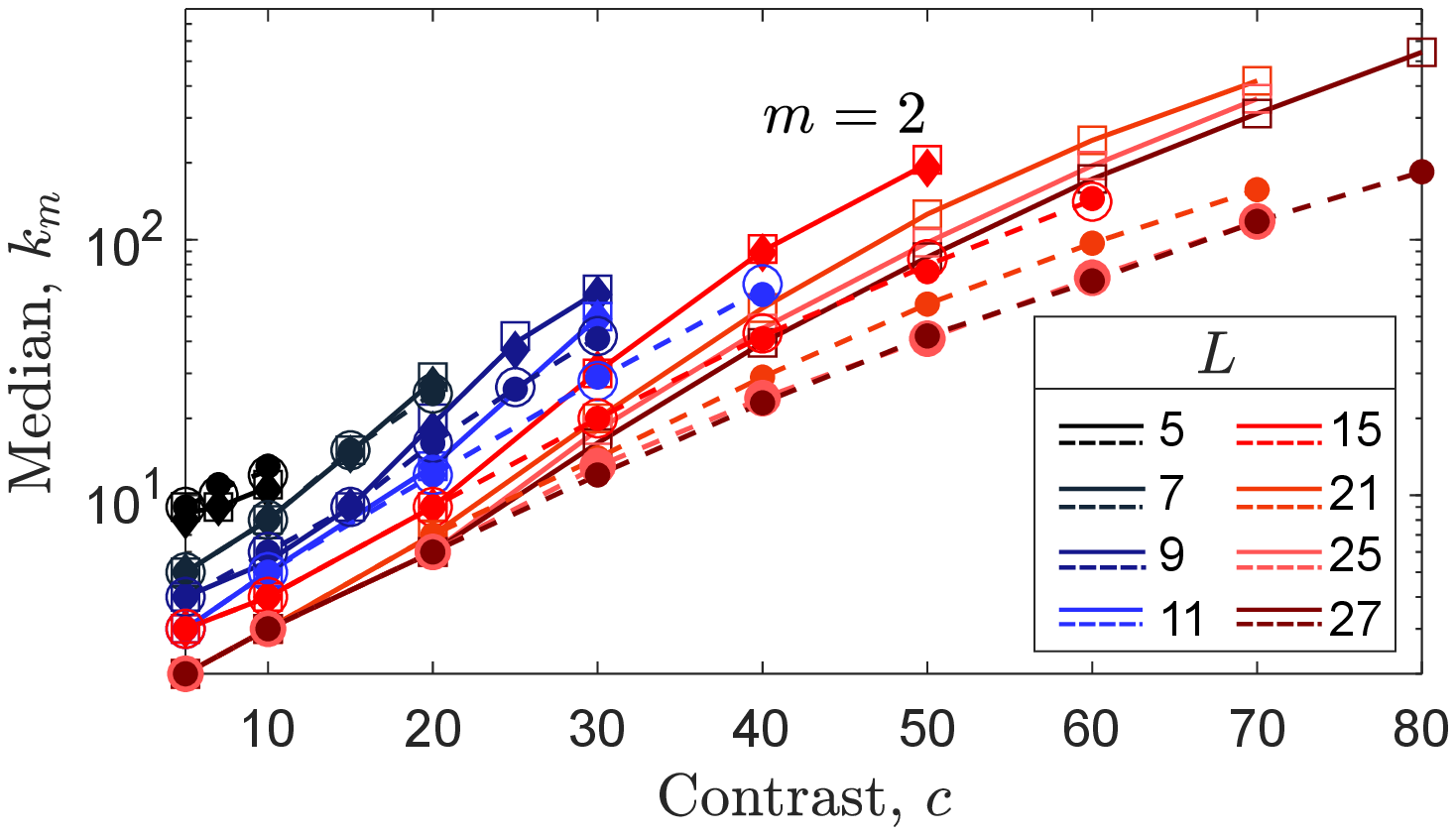}
(a)\\
\\
\includegraphics[width= \columnwidth]
{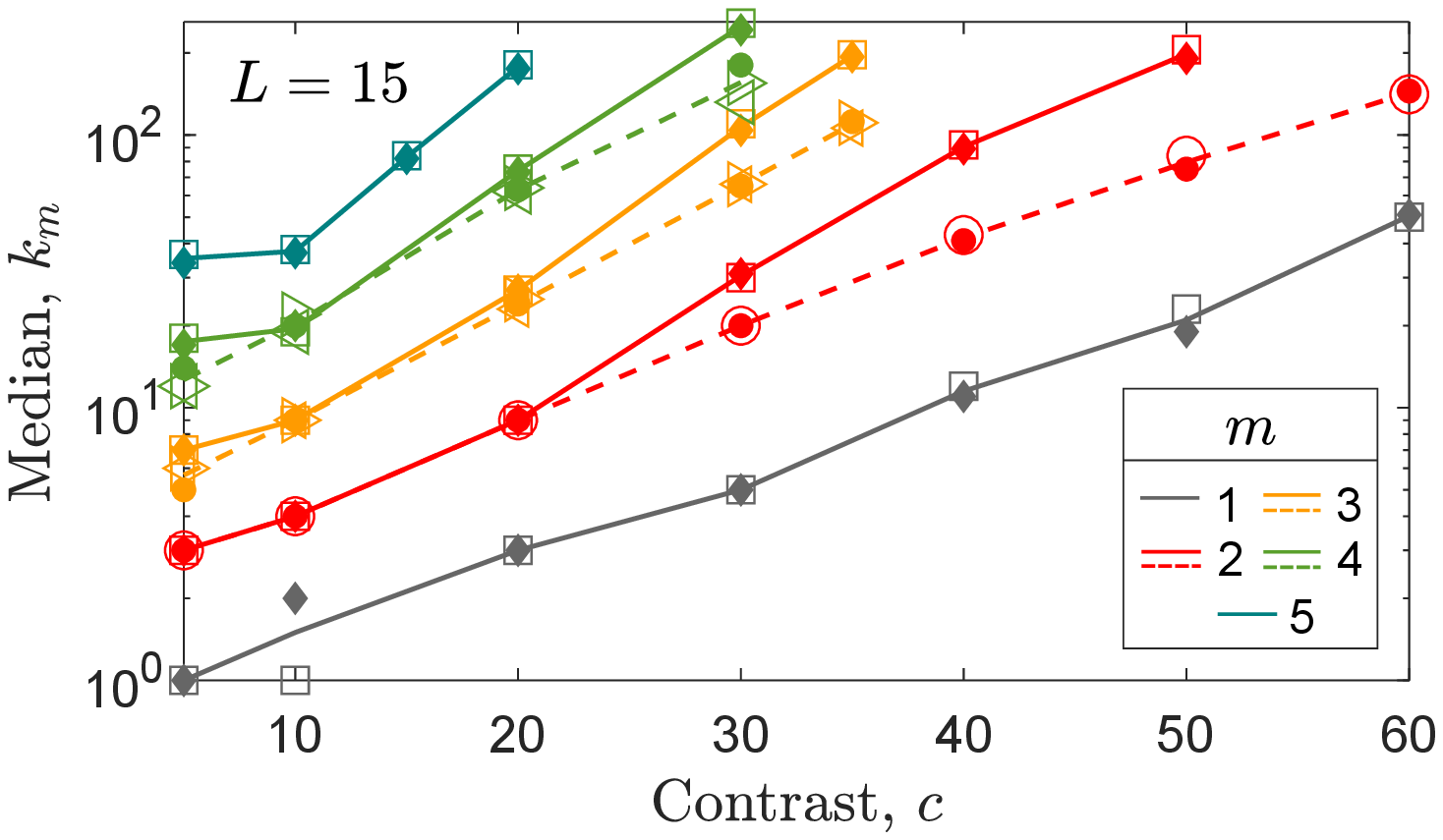}
\\
(b)\\
\includegraphics[width= \columnwidth]
{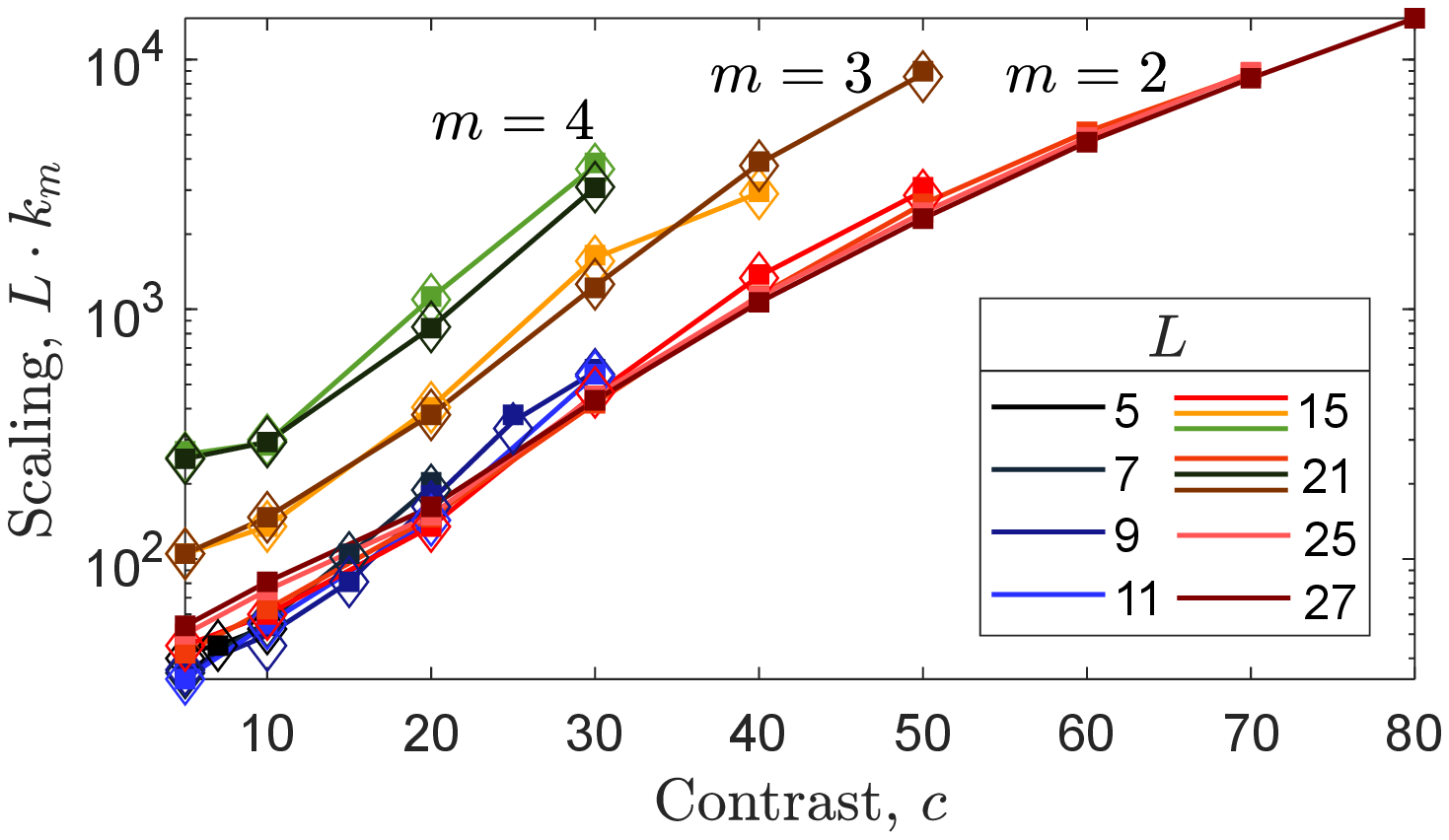}
\\
(c)
\end{tabular}
\caption{(color online)  (a) Network size $L$ with $m=2$ vs. contrast $c$ for the patterns `10' ($\bullet$), `01' ($\bigcirc$), `11' ($\blacklozenge$), and `00' ($\Box$). The lines are to guide an eye for exploring  differences between mixed `10', `01' and pure `00', `11' patterns. (b)  Outputs $m$ vs. contrast $c$ for networks of ${L=15}$ with patterns indicated in (a) and, additionally for `100', `1000' ($\bullet$); `101', `0110' ($\vartriangleleft$); `110', `1100' ($\vartriangleright$); `1', `111', `1111' ($\blacklozenge$), and `0', `000', `0000' ($\Box$).  (c) The data shown in (a) and (b) scaled as $L\cdot{}k_m$.}
	\label{fig:4}
\end{figure}

{\em Scaling.---}
Finally, to understand what controls the median number of pruned bonds required to reach a target pattern, we fix the number of frequencies at $n=2$ and explore $k_m$ for a range of system sizes $L$,  number of target nodes $m$, contrasts $c$, distinct target patterns, and frequency gaps.
First, we found that the gap between excitation frequencies can be varied over a wide range without affecting $k_m$, as long as $f_2-f_1 \gtrsim 0.006$, which is of the order of the eigenfrequency spacing -- hence correlations between the effect of pruning on different modes are essentially independent of the frequency ratio  (Supplemental Information).
Second, we investigate the effect of $L, m$ and $c$, for two fixed frequencies $f_1=1$ and $f_2=1.2$, and two pairs of target patterns (``switching'': `01' and `10', and ``on/off'': `00' and `11')~\cite{footnote1}. We find that $k_m$ increases with both contrast $c$ and $m$, as expected, yet decreases with $L$ (Fig.~\ref{fig:3}a-b). In addition, mixed patterns require less pruning, showing that pruning is a particularly promising strategy for acoustic steering.
Surprisingly, we can collapse the data for different sizes by
plotting $k_m\cdot{}L$ as function of $c$, for $m\geq{}2$
(Fig.~3c). This suggest that it is not the number of nodes, but rather the linear system size which controls the number of pruned bonds. 
Hence, $k_m$ is therefore not an extensive quantity.

{\em Summary and Outlook.--- } We have shown that vibrational properties of disordered networks can be controlled by rationally pruning a small number of bonds. The resulting structures do not show
discernible paths, but nevertheless allow steering acoustic energy to targeted nodes at pre-defined frequencies. Overall, our results demonstrate that pruning is a viable strategy to realize multi-objective optimization in disordered spring-based metamaterials. Finally, we observed that the number of bonds that are required to be pruned increases with more stringent design objectives such as
larger contrast or more output nodes, as expected. Surprisingly, for a given target functionality, the number of required pruned bonds increases linearly with system size $L$, so that the fraction of pruned bonds
diminishes with $L$. This suggests that larger systems would allow
 for even more functionalities to be integrated.

{\em Acknowledgements.} A.K. acknowledges the Center for Information Technology of the Faculty of Science and Engineering, University of Groningen for their support and for providing access to the Peregrine high performance computing cluster.

\bibliographystyle{apsrev}

\begin{thebibliography}{30}
\bibitem{footnote1}
We analyze 2500 systems for ${L=5-7}$; 1000 systems for ${L=7-15}$; 500 systems for ${L=21-27}$.
\end{thebibliography}

\end{document}


\newcommand{\beq}{\begin{equation}}
\newcommand{\eeq}{\end{equation}}

\title{Supplementary Material to \\
Multi-Frequency Acoustic Steering in Rationally Pruned Disordered Networks}

\author{Anastasiia O. Krushynska}
\affiliation{Engineering and Technology Institute Groningen (ENTEG), Faculty of Science and Engineering, University of Groningen, Nijenborgh 4, 9747 AG Groningen, The Netherlands}
\author{Martin van Hecke}
\affiliation{AMOLF, Science Park 104, 1098 XG Amsterdam, The Netherlands}\affiliation{Huygens-Kamerlingh Onnes Lab, Leiden University, PObox 9504, 2300 RA Leiden, The Netherlands}

\date{\today}

\maketitle

\begin{figure}[htb]
\includegraphics[scale=0.5]{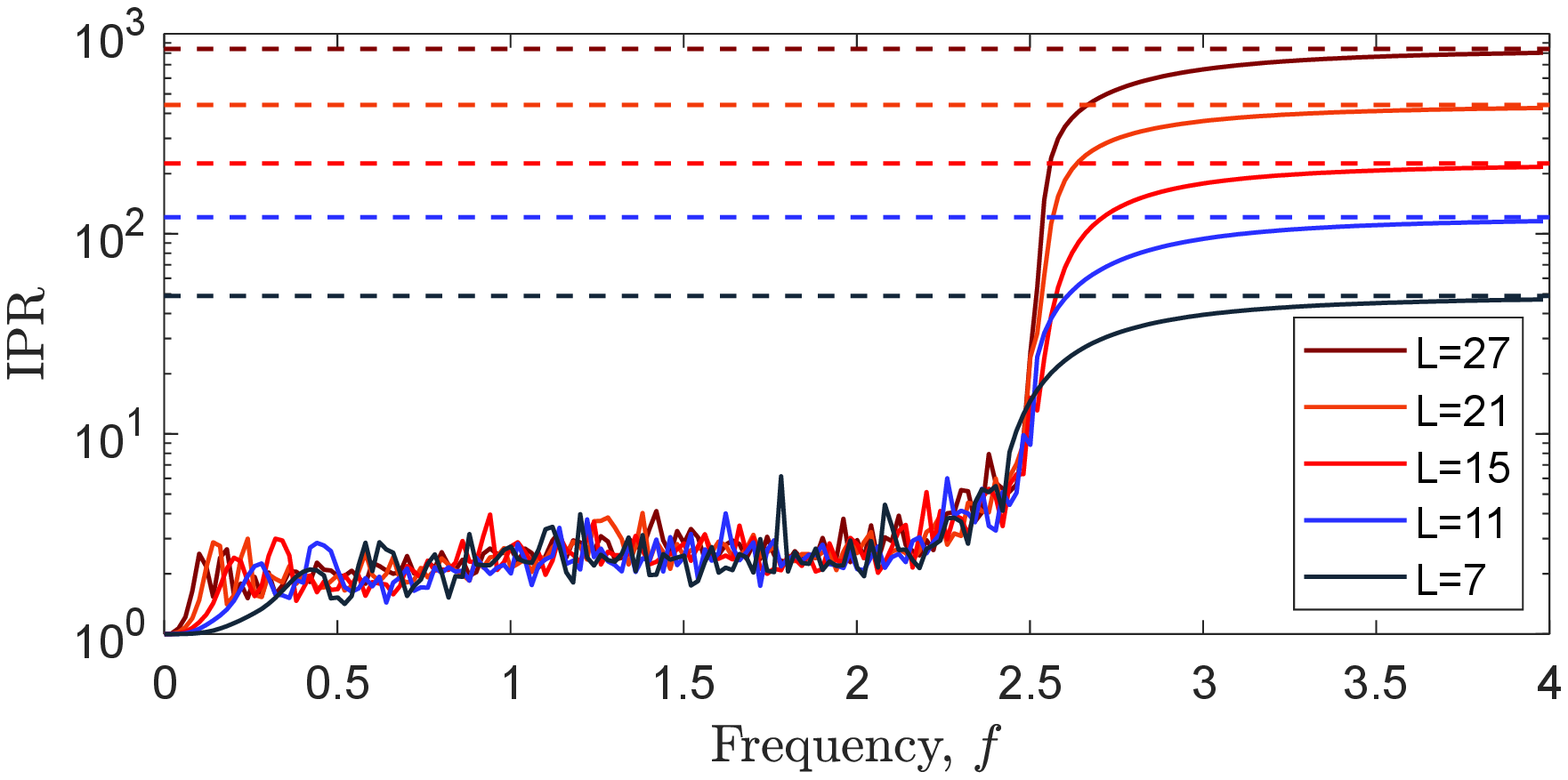}
	\caption{(color online) The inverse
participation ratio for output amplitudes at the target nodes $IPR:=<A_i^4>/<A_i^2>^2$ indicates spatially homogeneous vibration field showing crossover at $f=2.5$. At higher frequencies, strong mode localization occurs when the vibrational energy is concentrated on a single node as indicated by IPR tending to $O(N)$.}
	\label{fig:SM1}
\end{figure}

{\em IPR.---} To characterize the spatial distribution of the vibration modes, we have calculated the inverse participation ratio (IPR):
\begin{equation}
\mathrm{IPR} = \frac{< A_i^4>}{< A_i^2>^2},
\label{SI:eq_ipr}
\end{equation}
where $A_i$ are the $N$ amplitudes at each node, and
the $\mathrm{IPR}$ can range from 1 for a completely homogeneous mode to $N$ for a completely localized state. Our data shows that our modes become strongly localized at a size independent frequency around 2.5; the transition to extended modes occurs at a size dependent frequency lower than 0.5 (Fig.~\ref{fig:SM1}).

\begin{figure}[htb]
\includegraphics[width= 0.95\columnwidth]
{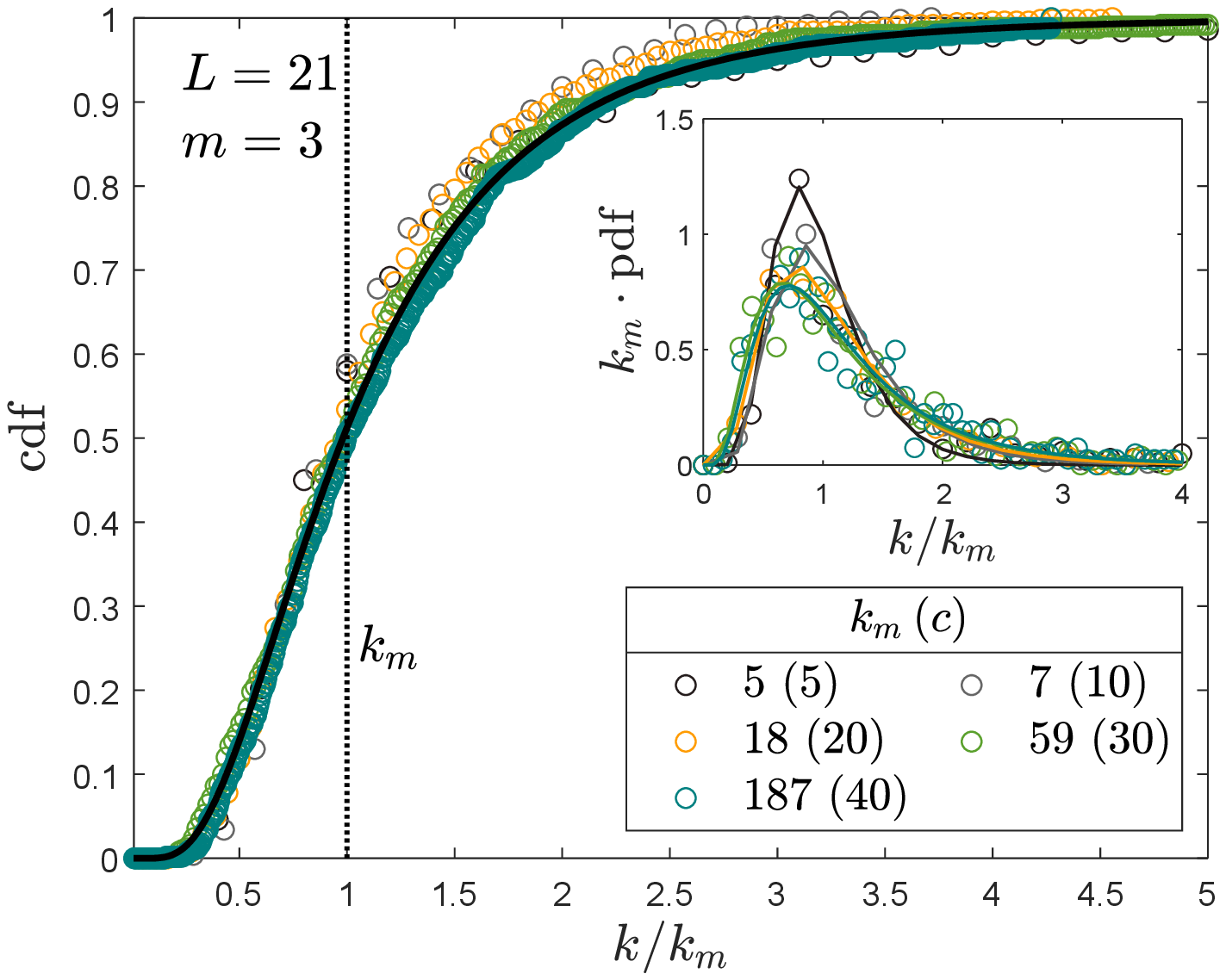}
	\caption{(color online) The cumulative distribution function (CDF) and the probability density function (pdf) for 500 systems of size ${L=21}$, $m=3$ with patterns `000' at $f_1=1$ and `111' at $f_2=1.2$ and different contrast values; the scaled median is indicated by the dotted line.  The data is perfectly fit by a log-normal distribution (the solid curve).
 }
	\label{fig:SM2}
\end{figure}

{\em Log-Normal distribution.---} {The CDFs of the number of cut bonds, $k$, show minor deviations
from a log-normal distribution --- in particular, when the long tail of the distribution of $k$ gets truncated for small systems, larger contrast, or more challenging patterns (see data for $L=15$ in the main text). Here we show that for larger system sizes ($L=21$) the CDFs of $k/k_m$ for $c=$5, 10, 20, 30, and 40 are very well matched by log-normal fits, provided that $k$ remains small (Fig.~\ref{fig:SM2}). }

\begin{figure*}[htb]
\begin{tabular}{cc}
\includegraphics[width=0.5\columnwidth]{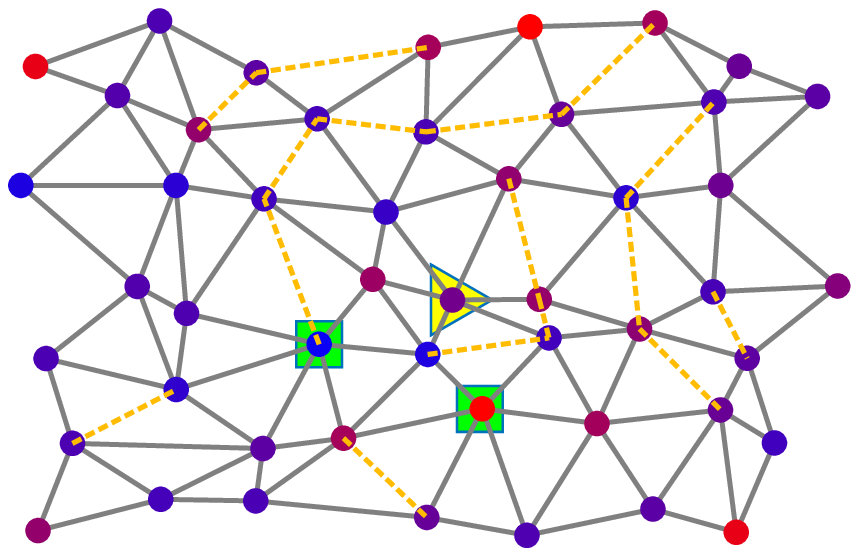}&
\includegraphics[width=0.55\columnwidth]{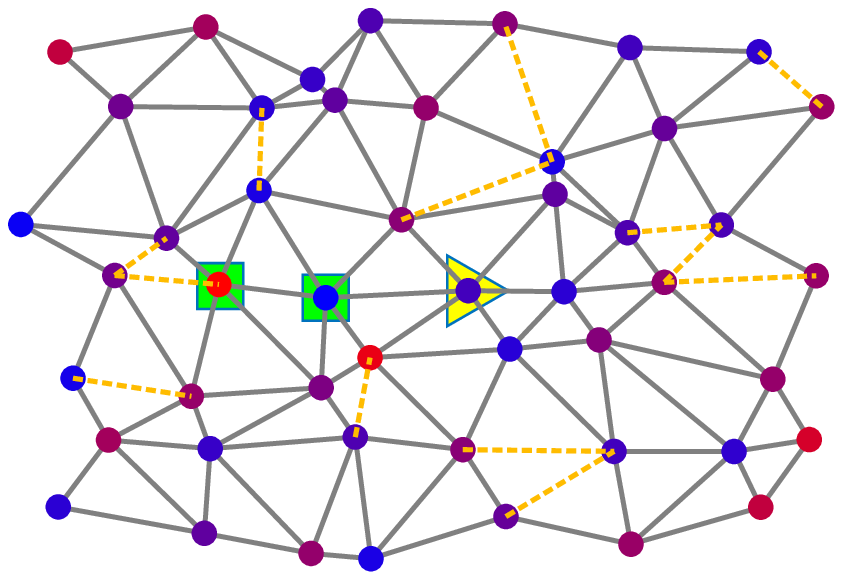}\\
\multicolumn{2}{c}{(a) $L=7$, $m=2$, $c=20$}\\
\includegraphics[width=0.6\columnwidth]{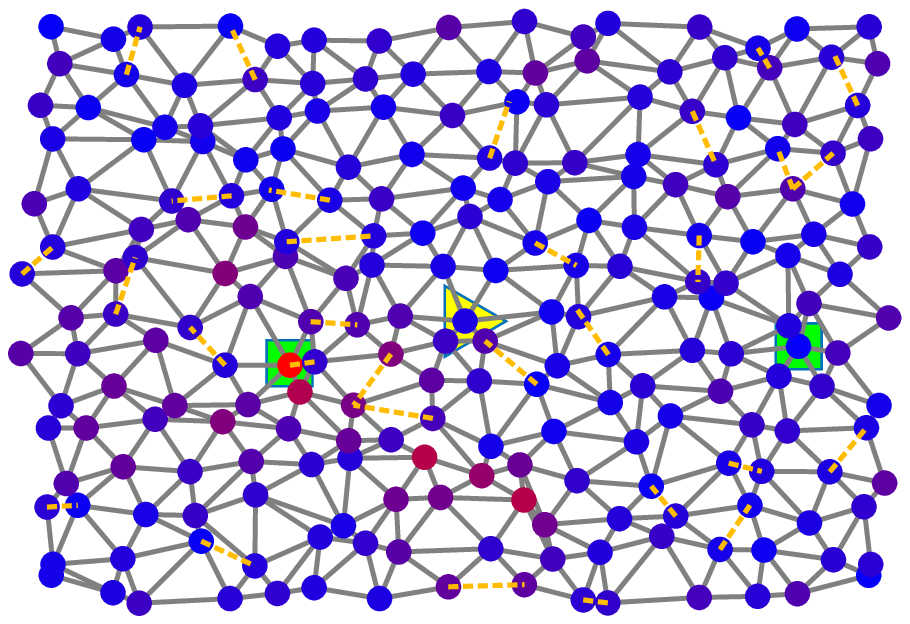}&
\includegraphics[width=0.6\columnwidth]{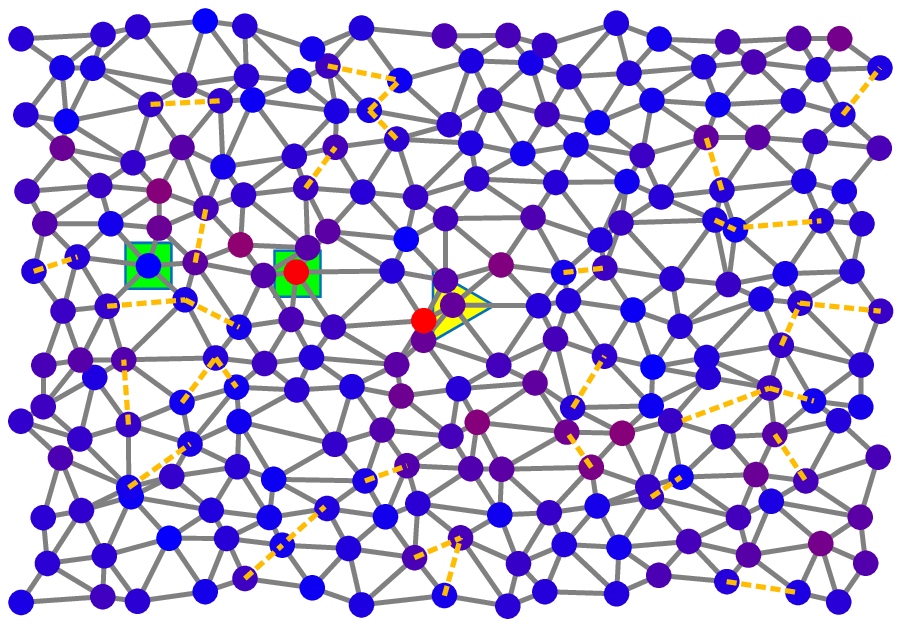}\\
\multicolumn{2}{c}{(b) $L=15$, $m=2$, $c=40$}\\
\multicolumn{2}{c}{\includegraphics[width=1.5\columnwidth]{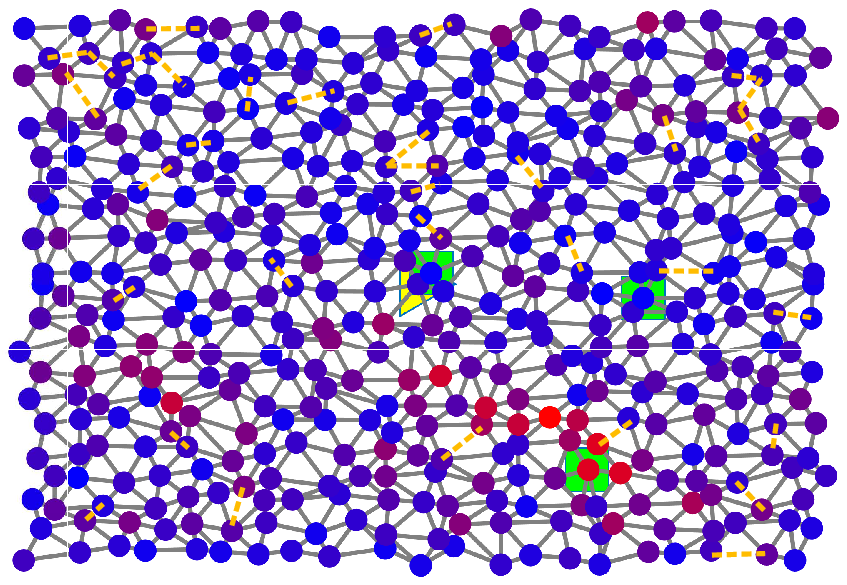}}\\
\multicolumn{2}{c}{(c) $L=21$, $m=3$, $c=30$}
\end{tabular}
	\caption{(color online) (a)-(c) Pruned bonds (orange dashed) are randomly spread in a network and seem to be independent of the positions of outputs, system size $L$ and the number of outputs $m$. The plotted data are calculated for the mixed patterns 01 at $f_1=1$ and 10 at $f_2=1.2$ (only the configurations for $f_1$ are shown).
 }
	\label{fig:SM3}
\end{figure*}

{\em Absence of real space pathways.---}
Examples of the locations of pruned bonds in various networks reveal no visible spatial correlations to the positions of source or targets (Fig.~\ref{fig:SM3}).

\begin{figure*}[t!]
\includegraphics[width=2\columnwidth]
{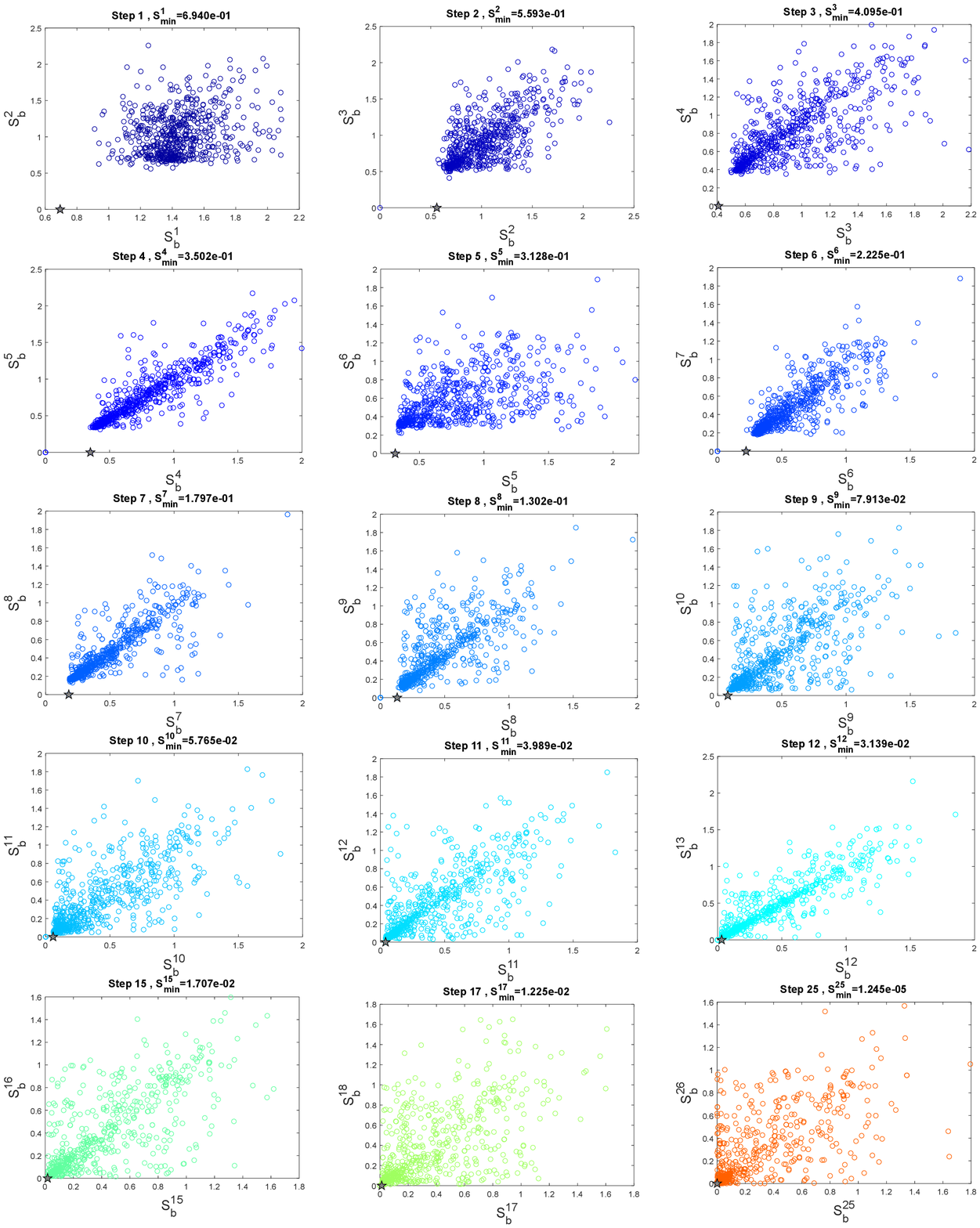}
\caption{Scatter plots of
$S_b^{k+1}$ vs $S_b^{k}$ (for non-pruned springs) for step-by-step variation of $k$ required to reach $c=40$ for mixed patterns `10' at $f_1=1$ and `01' at $f_2=1.2$ in a network of size $L=15$. We see
that the correlations are build and destroyed, thus necessitating a costly recalculation of the dynamical response at each pruning step. {The points corresponding to pruned bonds are indicated by the stars.}
}	\label{fig:SM4}
\end{figure*}

{\em Scatter plots.---} To illustrate the intermittent appearance and disappearance of correlations between subsequently pruned bonds, we present scatter plots of $S^{k+1}_b$ vs. $S^{k}_b$ for all values of $k$, for the same data as shown in Fig.~2 of the main text in Figure~\ref{fig:SM4}.

\begin{figure}[t!]
\begin{tabular}{c}
\includegraphics[scale=0.5]{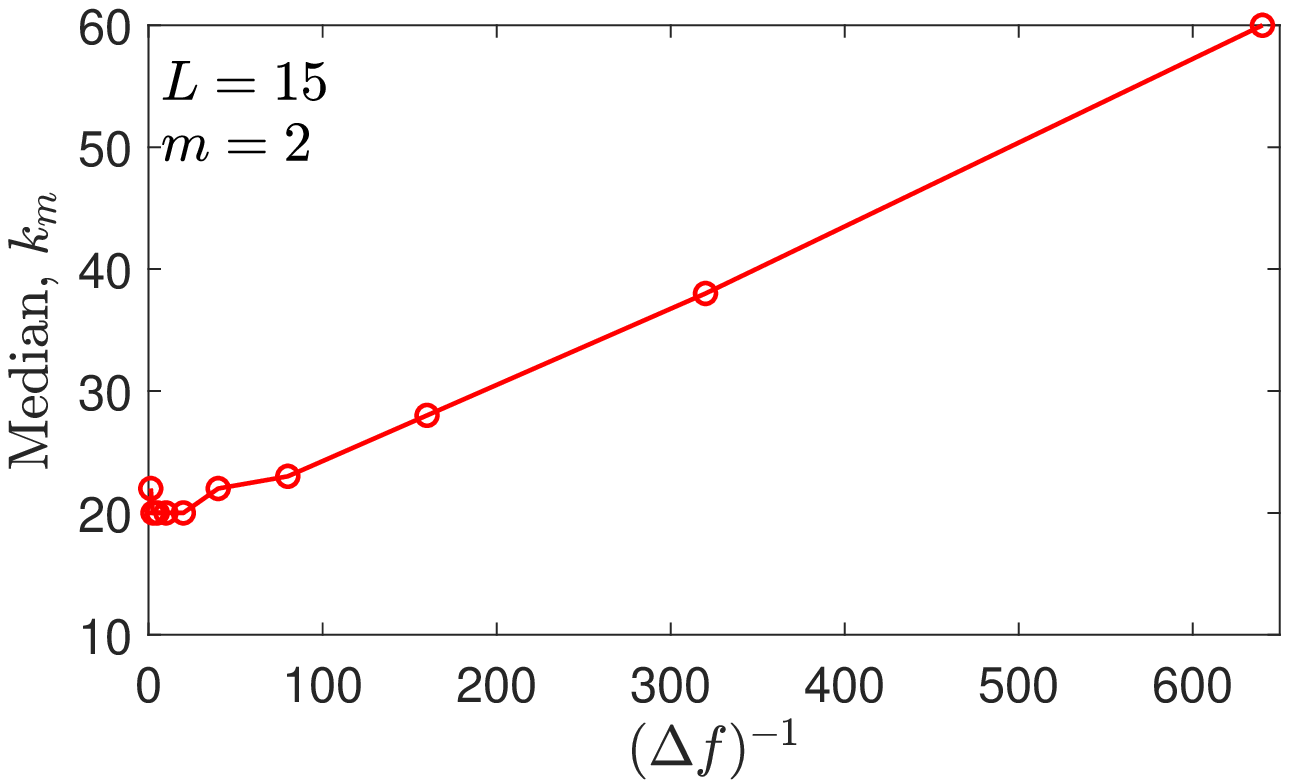}\\
(a)\\
\includegraphics[scale=0.5]{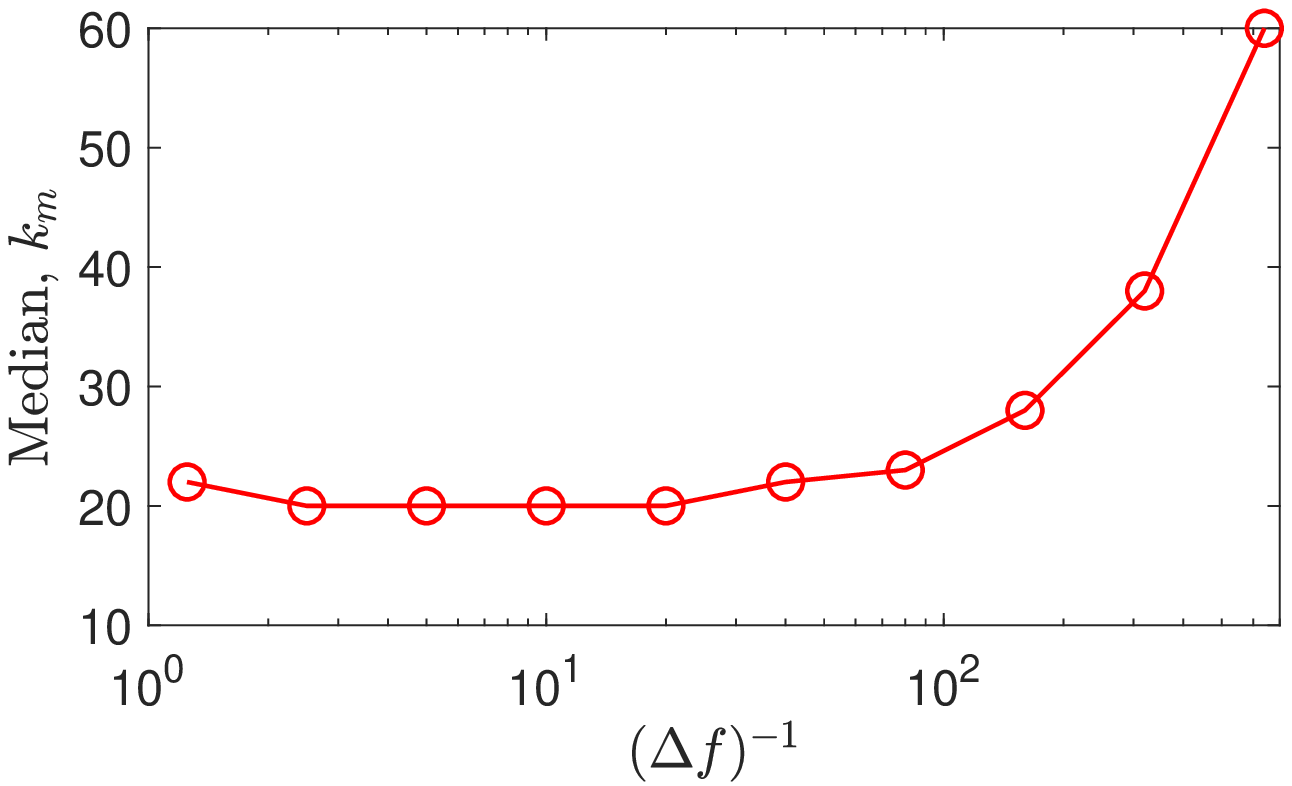}\\
(b)
\end{tabular}
	\caption{(color online) Median {\em vs.} varying frequency ${f_2=f_1+\Delta f}$ for 1000 systems of size ${L=15}$ with ${m=2}$ and mixed patterns `01' at $f_1$, `10' at $f_2$. The difference in eigenfrequencies for such systems is around 0.0055.
 }
	\label{fig:SM5}
\end{figure}

{\em Frequency spacing.---} We illustrate the median $k_m$ for a given set of targets as function of the frequency spacing $\Delta f:= f_2-f_1$ in Fig.~\ref{fig:SM5}. Essentially, $k_m$ converges to a well defined value for large frequency differences, and only increases significantly when $\Delta f$ falls below (0.01).

{\em Data.---} In tables~\ref{table:SM1}-\ref{table:SM4} we provide quantitative details for the data shown in Fig.~4 of the main text.

\begin{table*}[bp]
\caption{\label{table:SM1} Medians of the pruned bonds (in \% to the total number of bonds in a network) for two random outputs, $m=2$; failed cases are excluded from the statistics. No number means the absence of convergence; `NA' refers to Not Analyzed cases.}
\begin{tabular}{ l | l | r r r r ||  l | l | r r r r}
\hline
\hline
Size&Contrast&\multicolumn{4}{c||}{Pattern}&Size&Contrast&\multicolumn{4}{c}{Pattern}\\
\hline
$L$&$c$& 10&01&00&11&$L$&$c$& 10&01&00&11\\
\hline
5&5&16.1\%&16.1\%&16.1\%&14.3\% & 21&5&0.16\%&NA&0.16\%&NA\\
&7&19.6\%&17.8\%&16.1\%&16.1\%  &  &10&0.24\%&NA&0.24\%&NA\\
&10&23.2\%&21.4\%&19.6\%&18.7\% &  &20&0.56\%&NA&0.56\%&NA\\
&&&&&                           &  &30&1.1\%&NA&1.6\%&NA\\
7&5&4.2\%&4.2\%&4.2\%&4.2\%     &  &40&2.3\%&NA&4.4\%&NA\\
&10&6.7\%&6.7\%&6.7\%&6.7\%     &  &50&4.5\%&NA&10.2\%&NA\\
&15&12.5\%&12.5\%&12.5\%&12\%   &  &60&7.8\%&NA&19.8\%&NA\\
&20&20.8\%&20.8\%&24.2\%&22.5\% &  &70&12.7\%&NA&-&NA\\
&&&&&\\
9&5&1.9\%&1.9\%&1.9\%&1.9\%     & 25&5&0.11\%&NA&0.11\%&NA\\
&10&2.9\%&2.9\%&2.9\%&2.4\%     &  &10&0.17\%&NA&0.17\%&NA\\
&15&4.3\%&4.3\%&4.3\%&4.3\%     &  &20&0.34\%&NA&0.34\%&NA\\
&20&7.7\%&7.7\%&9.6\%&8.6\%     &  &30&0.7\%&NA&1\%&NA\\
&25&12.5\%&12.7\%&20.2\%&17.8\% &  &40&1.3\%&NA&2.5\%&NA\\
&30&19.7\%&20.2\%&30.8\%&29.3\% &  &50&2.3\%&NA&5.5\%&NA\\
&&&&&                           &  &60&4\%&NA&11\%&NA\\
11&5&0.9\%&0.9\%&0.9\%&0.9\%    &  &70&4.3\%\%&NA&20\%&NA\\
&10&1.6\%&1.6\%&1.6\%&1.6\%     &  & & & &\\
&20&3.7\%&3.7\%&4\%&4\%         & 27&5&0\%&NA&0\%&NA\\
&30&9\%&8.7\%&15.6\%&15.3\%     &   &10&0.12\%&NA&0.12\%&NA\\
&40&19\%&20.9\%&-&-             &   &20&0.42\%&NA&0.42\%&NA\\
&&&&&                           &   &30&0.5\%&NA&0.67\%&NA\\
15&5&0.5\%&0.5\%&0.5\%&0.5\%    &   &40&0.95\%&NA&1.6\%&NA\\
&10&0.6\%&0.6\%&0.6\%&0.6\%     &   &50&1.7\%&NA&3.5\%&NA\\
&20&1.5\%&1.5\%&1.5\%&1.5\%     &   &60&2.9\%&NA&7.2\%&NA\\
&30&3.2\%&3.2\%&4.9\%&5\%       &   &70&4.9\%&NA&13\%&NA\\
&40&6.7\%&7\%&14.9\%&14.4\%     &   &80&7.7\%&NA&22.5\%&NA\\
&50&12.2\%&13.6\%&33.4\%&31\%\\
&60&23.5\%&22.9\%&-&-\\
\hline
\hline
\end{tabular}
\end{table*}

\begin{table}[ht]
\caption{\label{table:SM2} Medians of the pruned bonds (in \% to the total number of bonds in a network) for three random outputs, $m=3$ in 1000 (${L=15}$) and 500 (${L=21}$) networks; failed cases are excluded from the statistics. No number means the absence of convergence; `NA' refers to Not Analyzed cases.}
\begin{tabular}{ l | l | r r r r r}
\hline
\hline
Size&Contrast&\multicolumn{5}{c}{Pattern}\\
\hline
$L$&$c$& 110&100&101&000&111\\
\hline
15&5&1\%&0.8\%&1\%&1.1\%&1.1\%\\
  &10&1.5\%&1.5\%&1.5\%&1.5\%&1.5\%\\
  &20&4\%&3.9\%&3.7\%&4.4\%&4.4\%\\
  &30&10.7\%&10.6\%&10.7\%&17.8\%&16.9\%\\
  &35&18\%&18.2\%&17.2\%&32\%&31.4\%\\
\hline
21&5&NA&NA&NA&0.4\%&0.4\%\\
  &10&NA&NA&NA&0.6\%&0.6\%\\
  &20&NA&NA&NA&1.4\%&1.4\%\\
  &30&NA&NA&NA&4.7\%&4.8\%\\
  &40&NA&NA&NA&15\%&14.4\%\\
  &50&NA&NA&NA&34.5\%&32.8\%\\
\hline
\hline
\end{tabular}
\end{table}

\begin{table}[h]
\caption{\label{table:SM3} Medians of the pruned bonds (in \% to the total number of bonds in a network) for four random outputs, $m=4$, in 500 networks; failed cases are excluded from the statistics. `NA' indicates Not Analyzed cases.}
\begin{tabular}{ l | l | r r r r r}
\hline
\hline
Size&Contrast&\multicolumn{5}{c}{Pattern}\\
\hline
$L$&$c$& 1100&0110&1000&0000&1111\\
\hline
15& 5&1.9\%&1.9\%&2.3\%&2.9\%&2.8\%\\
  &10&3.6\%&3.1\%&3.2\%&3.1\%&3.2\%\\
  &20&10.4\%&10\%&10.2\%&12.2\%&11.8\%\\
  &30&35.6\%&21.4\%&29.3\%&41.7\%&39.4\%\\
\hline
21&5&NA&NA&NA&1\%&1\%\\
  &10&NA&NA&NA&1.1\%&1.1\%\\
  &20&NA&NA&NA&3.2\%&3.3\%\\
  &30&NA&NA&NA&11.8\%&11.5\%\\
  &40&NA&NA&NA&36.5\%&35.7\%\\
\hline
\hline
\end{tabular}
\end{table}

\begin{table}[h]
\caption{\label{table:SM4} Medians of the pruned bonds (in \% to the total number of bonds in a network) for five random outputs, $m=5$, in 500 networks of size $L=15$; failed cases are excluded.}
\begin{tabular}{ l | r r r}
\hline
\hline
Contrast&\multicolumn{2}{c}{Pattern}\\
\hline
$c$&0000&1111\\
\hline
 5&5.9\%&5.5\%\\
10&6.2\%&6\%\\
15&13.6\%&13.3\%\\
20&29.4\%&28\%\\
\hline
\hline
\end{tabular}
\end{table}